\documentstyle[psfig]{mn}

\newif\ifAMStwofonts

\ifoldfss
 \ifCUPmtlplainloaded \else
 \NewTextAlphabet{textbfit} {cmbxti10} {}
 \NewTextAlphabet{textbfss} {cmssbx10} {}
 \NewMathAlphabet{mathbfit} {cmbxti10} {} 
 \NewMathAlphabet{mathbfss} {cmssbx10} {} 
 \fi
 \ifAMStwofonts
 \ifCUPmtlplainloaded \else
 \NewSymbolFont{upmath} {eurm10}
 \NewSymbolFont{AMSa} {msam10}
 \NewMathSymbol{\upi} {0}{upmath}{19}
 \NewMathSymbol{\umu} {0}{upmath}{16}
 \NewMathSymbol{\upartial}{0}{upmath}{40}
 \NewMathSymbol{\leqslant}{3}{AMSa}{36}
 \NewMathSymbol{\geqslant}{3}{AMSa}{3E}

  \let\le=\leqslant
  
 \fi
 \fi
\fi 

\ifnfssone
 \newmathalphabet{\mathit}
 \addtoversion{normal}{\mathit}{cmr}{m}{it}
 \addtoversion{bold}{\mathit}{cmr}{bx}{it}
 \newmathalphabet{\mathbfit} 
 \addtoversion{normal}{\mathbfit}{cmr}{bx}{it}
 \addtoversion{bold}{\mathbfit}{cmr}{bx}{it}
 \newmathalphabet{\mathbfss} 
 \addtoversion{normal}{\mathbfss}{cmss}{bx}{n}
 \addtoversion{bold}{\mathbfss}{cmss}{bx}{n}
 \ifAMStwofonts
 \ifCUPmtlplainloaded \else
 \UseAMStwoboldmath
 \makeatletter
 \new@mathgroup\upmath@group
 \define@mathgroup\mv@normal\upmath@group{eur}{m}{n}
 \define@mathgroup\mv@bold\upmath@group{eur}{b}{n}
 \edef\UPM{\hexnumber\upmath@group}
 \new@mathgroup\amsa@group
 \define@mathgroup\mv@normal\amsa@group{msa}{m}{n}
 \define@mathgroup\mv@bold\amsa@group{msa}{m}{n}
 \edef\AMSa{\hexnumber\amsa@group}
 \makeatother
 \mathchardef\upi="0\UPM19
 \mathchardef\umu="0\UPM16
 \mathchardef\upartial="0\UPM40
 \mathchardef\leqslant="3\AMSa36
 \mathchardef\geqslant="3\AMSa3E

  \let\le=\leqslant

 \fi
 \fi
\fi 

\ifnfsstwo
 \DeclareMathAlphabet{\mathbfit}{OT1}{cmr}{bx}{it}
 \SetMathAlphabet\mathbfit{bold}{OT1}{cmr}{bx}{it}
 \DeclareMathAlphabet{\mathbfss}{OT1}{cmss}{bx}{n}
 \SetMathAlphabet\mathbfss{bold}{OT1}{cmss}{bx}{n}
 \ifAMStwofonts
 \ifCUPmtlplainloaded \else
 \DeclareSymbolFont{UPM}{U}{eur}{m}{n}
 \SetSymbolFont{UPM}{bold}{U}{eur}{b}{n}
 \DeclareSymbolFont{AMSa}{U}{msa}{m}{n}
 \DeclareMathSymbol{\upi}{0}{UPM}{"19}
 \DeclareMathSymbol{\umu}{0}{UPM}{"16}
 \DeclareMathSymbol{\upartial}{0}{UPM}{"40}
 \DeclareMathSymbol{\leqslant}{3}{AMSa}{"36}
 \DeclareMathSymbol{\geqslant}{3}{AMSa}{"3E}

  \let\le=\leqslant

 \fi
 \fi
\fi 

\ifCUPmtlplainloaded \else
 \ifAMStwofonts \else 
 \def\upi{\pi}
 \def\umu{\mu}
 \def\upartial{\partial}
 \fi
\fi

\begin{document}

\title{Discovery of unusual pulsations in the cool, evolved Am stars HD~98851 and HD~102480}
\author[Santosh Joshi et al.]
{Santosh Joshi$^{1}$,\thanks{E-mail: santosh@upso.ernet.in} V. Girish$^{2}$, 
R. Sagar$^{1}$, D. W. Kurtz$^{3}$, P. Martinez$^{4}$, B. Kumar$^{1},$\\
\\
\LARGE {S. Seetha$^{2}$, B. N. Ashoka$^{2}$ and A. ~Zhou$^{5}$}\\
\\
$^{1}$State Observatory, Manora Peak, Naini Tal 263\,129, India\\
$^{2}$ISRO Satellite Centre, Airport Road, Bangalore 560\,034, India\\
$^{3}$Centre for Astrophysics, University of Central Lancashire, 
Preston PR1 2HE, UK\\
$^{4}$South African Astronomical Observatory, P.O. Box 9, Observatory 7935, South Africa\\
$^{5}$National Astronomical Observatories, Chinese Academy of Sciences, Beijing 
100\,012, P. R. China}

\date{Accepted ---------.
 Received ---------;
} 

\pagerange{\pageref{firstpage}--\pageref{lastpage}}
\pubyear{2001}

\maketitle

\label{firstpage}

\begin{abstract}
The chemically peculiar (CP) stars HD\,98851 and HD\,102480 have been
discovered to be unusual pulsators during the ``Naini Tal Cape Survey''
programme to search for pulsational variability in CP stars. Time series
photometric and spectroscopic observations of these newly discovered stars
are reported here. Fourier analyses of the time series photometry reveal
that HD\,98851 is pulsating mainly with frequencies 0.208\,mHz and 0.103\,mHz, 
and HD\,102480 is pulsating with frequencies 0.107\,mHz, 0.156\,mHz  
and 0.198 mHz. The frequency identifications are all subject to 1~d$^{-1}$
cycle count ambiguities. We
have matched the observed low resolution spectra of HD\,98851 and HD\,102480
in the range 3500-7400\,\AA  ~with theoretical synthetic spectra using
Kurucz models with solar metallicity and a micro-turbulent velocity 2
km\,s$^{-1}$. These yield $T_{eff}=7000\pm250$\,K, log\,$g=3.5 \pm 0.5$ for
HD\,98851 and $T_{eff} = 6750 \pm 250$\,K, log\,$g = 3.0 \pm 0.5$ for
HD\,102480. We determined the equivalent H-line spectral class of these 
stars to be
F1\,IV and F3\,III/IV, respectively. A comparison of the location of
HD\,98851 and HD\,102480 in the HR diagram with theoretical stellar
evolutionary tracks indicates that both stars are about 1-Gyr-old,
2-$M_{\odot}$ stars that lie towards the red edge of the $\delta$ Sct
instability strip. From comparison between the observed and calculated
physical parameters, we conclude that HD\,98851 and HD\,102480 are cool,
evolved Am pulsators. The light curves of these pulsating stars have 
alternating high and low amplitudes, nearly harmonic (or sub-harmonic) 
period ratios, high pulsational overtones and Am spectral types. This is unusual 
for both Am and $\delta$ Sct pulsators, making these stars 
interesting objects for further observational and theoretical studies. 
\end{abstract}

\begin{table*}
\caption{Data for HD\,98851 and HD\,102480 taken from Hauck \& Mermilliod (1998), 
Abt (1984) and ESA (1997).}
\begin{center}
\begin{tabular}{cccccccccc}
\hline
&& \\
Star & $\alpha_{2000} $ & $\delta_{2000}$ &$m_v$ & $B-V$ & $V-I$&
$b-y$ & $m_1$ & $c_1$ & Sp. Type based on\\
 & & &(mag) & (mag) & (mag) & (mag) & & & K/H/Metal lines \\
&& \\
\hline
&& \\
HD\,98851 & $ 11^h 22^m 48^s $ & $ 31^\circ 49^{'} 32^{''}$ & 7.40 & 
$ 0.33\pm0.00 $ & $ 0.39\pm0.01$ & 0.199 & 0.222 & 0.766 & 
F1/F1IV/F3 \\
&&&&&&& \\
HD\,102480 & $ 11^h 47^m 52^s $ & 
$ 53^\circ 00^{'}54^{''}$ & 8.40 & $ 0.36\pm 0.01 $ & 
$ 0.42\pm 0.02 $ & 0.211 & 0.204 & 0.732 & F2/F4/F4 \\
&& \\
\hline
\end{tabular}
\end{center}
\end{table*} 

\begin{figure*}
\hbox{
\psfig{height=3.5in,width=3.5in,file=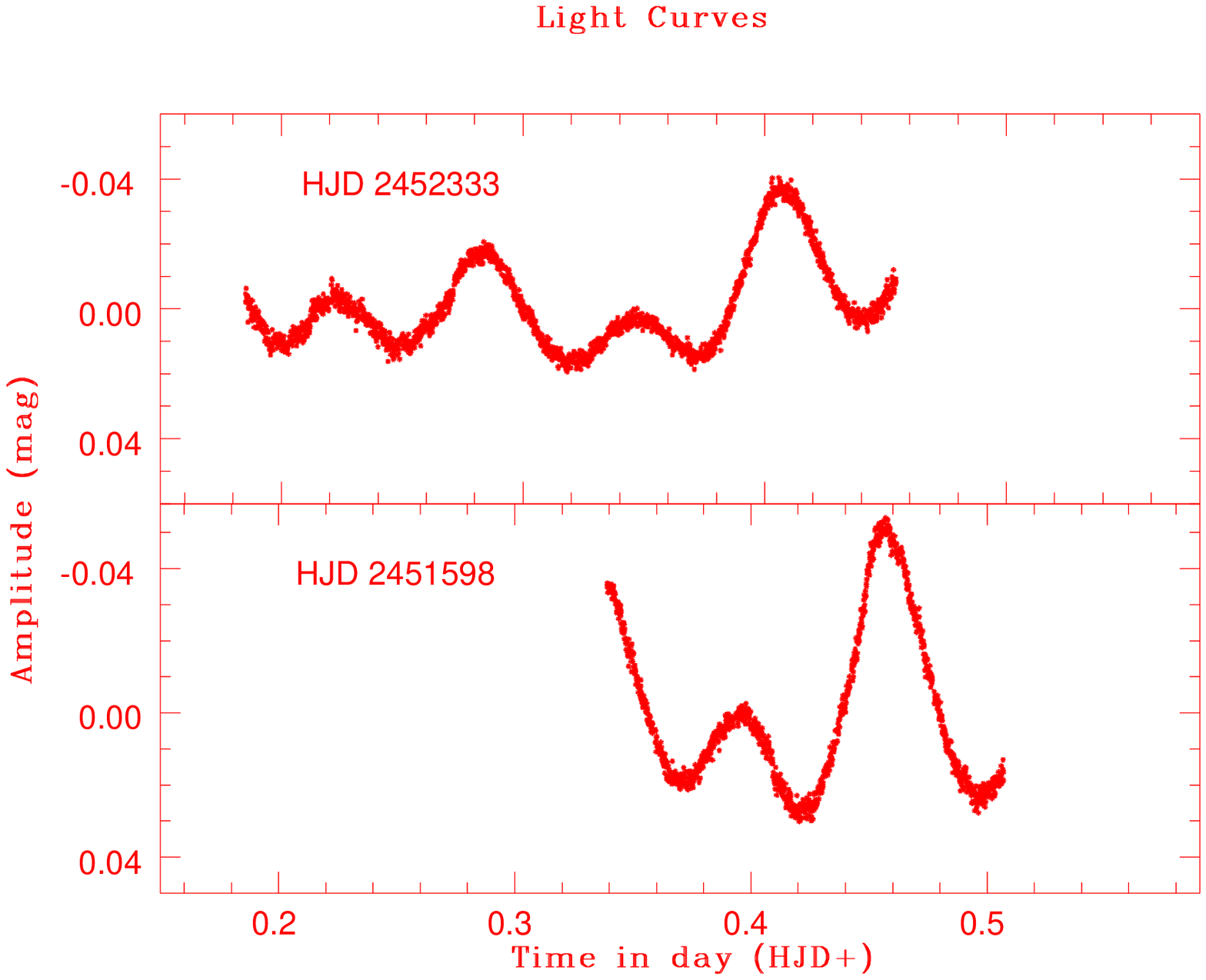}
\psfig{height=3.5in,width=3.5in,file=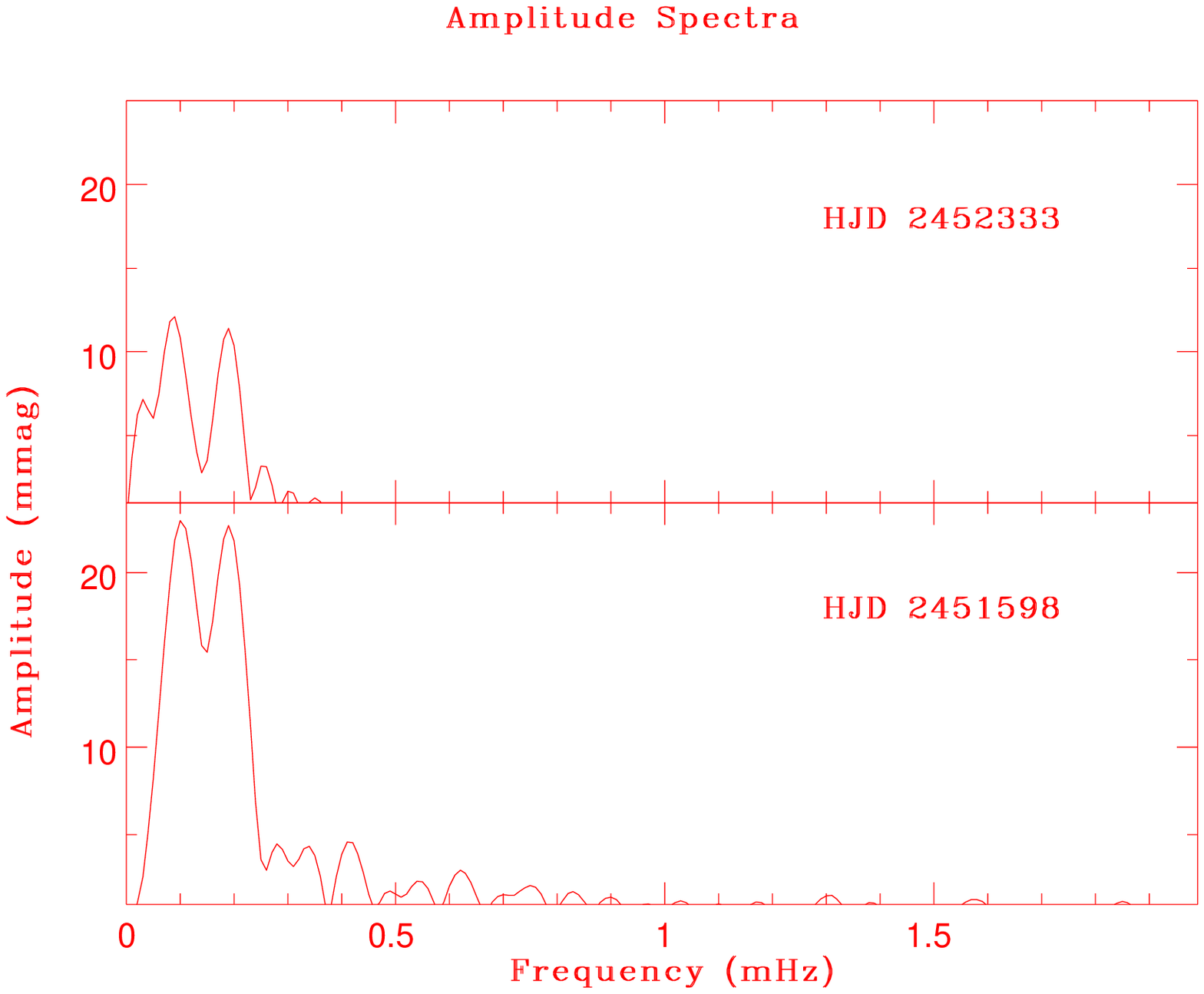}
}
\vspace{-0.5in}
\caption{The Johnson $B$ light curves of HD\,102480 and their corresponding 
amplitude spectra. The upper panels are for 27 Feb 2002 (HJD 2452333) and 
the lower panels for 23 Feb 2000 (HJD 2451598). Both light curves were
taken at Naini Tal. The amplitude spectra show two prominent peaks at 0.09\,mHz and 0.19\,mHz.}
\end{figure*} 

\section{Introduction}

The chemically peculiar (CP) stars lying in the temperature range
7400\,K\,$\le T_{eff} \le$\,10,200\,K are identified by the presence of
anomalously strong (and/or weak) absorption lines of certain heavy and rare
earth elements in their spectra. Two main classes of peculiarities are
recognised among CP stars in the spectral range B5 to F5. The first class
comprises
the peculiar A type stars, designated as `Ap', whose spectra are characterised by
over-abundances of Si, Mn, Cr, Sr and rare earth elements. The other class
comprises the
metallic line stars, designated as `Am' stars, whose spectra are characterised
by an underabundance of Ca (and/or Sc) and/or an overabundance of the Fe
group and heavier elements with respect to normal stars of same colour. Am stars with spectral types based on metallic lines and the Ca\,II K-line differing by five or more sub-classes are called classical Am stars; otherwise they are called marginal Am stars (designated as `Am:').

About 30\% of the stars in the lower classical instability strip with luminosities
ranging from the zero age main sequence (ZAMS) to about 2 mag above the main
sequence, with spectral types in the range late A to early F, and with
masses of 1.5\,M$_{\odot}$ to 2.5\,M$_{\odot}$ are $\delta$ Sct stars. These
stars pulsate in radial and/or non-radial $p$ (and possibly $g$) modes
driven by the $\kappa $ mechanism, with periods between 30 min and 8 hr and
amplitudes varying from a few mmag to almost a magnitude.

In the spectral range A to F almost 70\% of non-CP stars are $\delta$ Sct
variables at the current level of sensitivity. Most non-variables are Am
stars. Therefore the discovery of $\delta$ Sct pulsations in Am stars is
very important because this type of pulsation and anomalous abundance are
known to coexist in very few stars.  Among the CP stars which exhibit
$\delta$ Sct pulsations are several luminous, cool, evolved Am stars. The
pulsations in evolved Am stars and marginal Am stars lying near the red edge
of the instability strip are accounted for within the diffusion hypothesis;
ZAMS Am stars are predicted not to pulsate (Turcotte et al. 2000).  For a
recent review of the different classes of CP stars and a discussion of
pulsation in the presence of metallicism, the reader is referred to Kurtz \&
Martinez (2000), Kurtz (2000) and references therein.

\begin{figure*}
\hbox{
\psfig{height=5.5in,width=3.666in,file=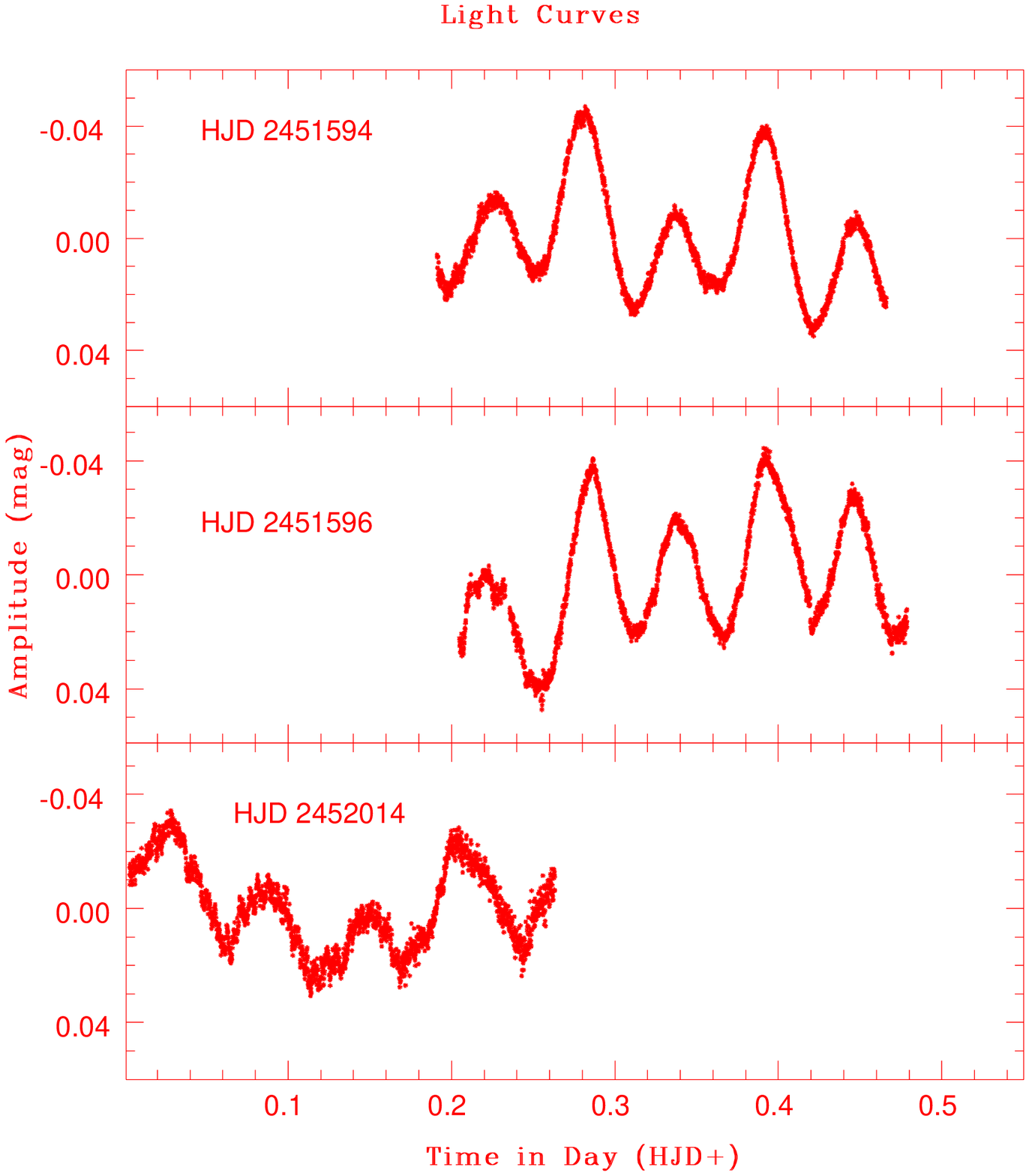}
\psfig{height=5.5in,width=3.666in,file=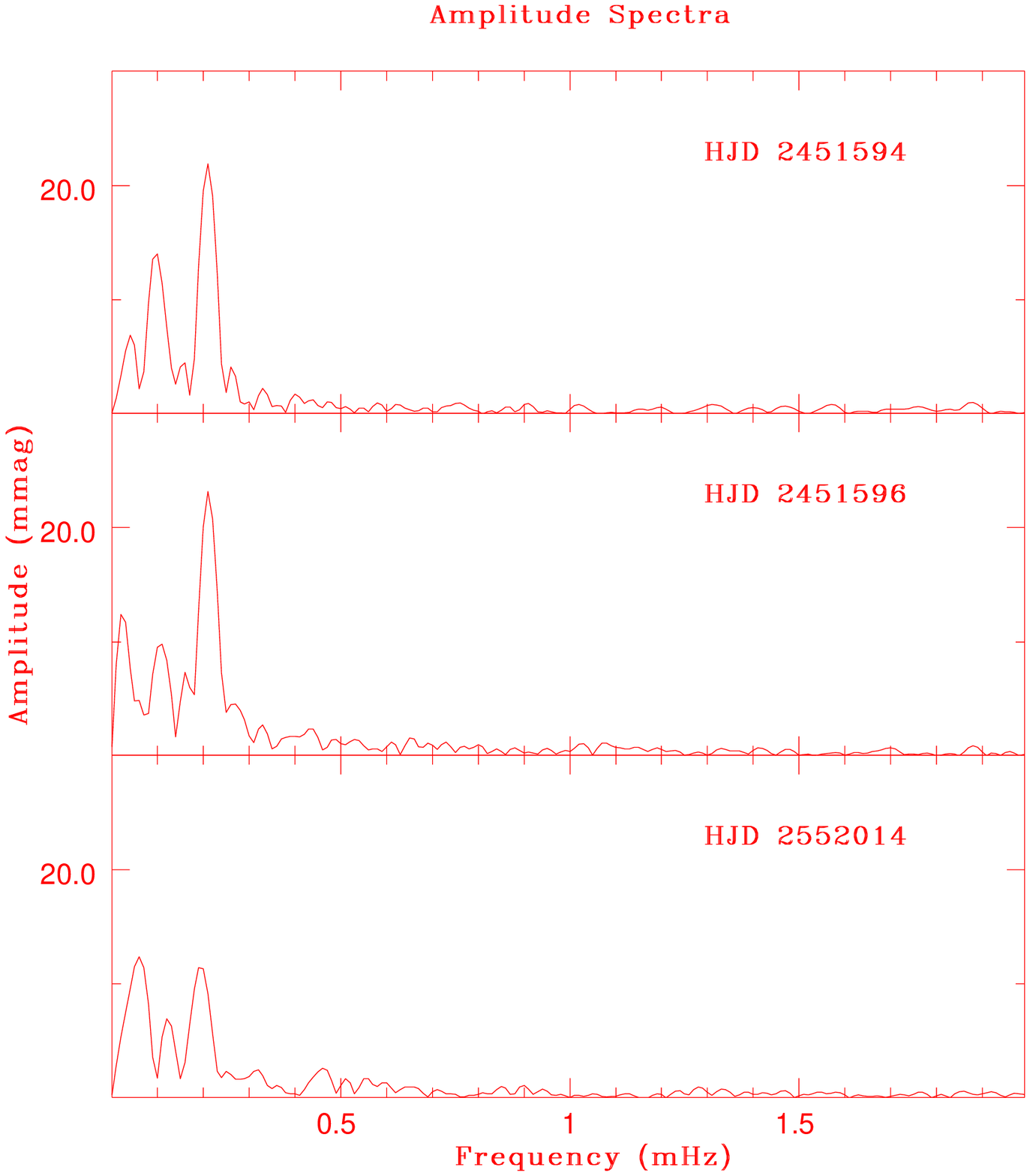}
}
\caption{Typical light curves of HD\,98851 and their corresponding amplitude
spectra. The top two panels are the
light curves obtained in Johnson $B$ from Naini Tal. The third panel shows the Johnson $V$ data from Xinglong.}
\end{figure*} 

To search for pulsations in the CP stars, we started the ``Naini Tal-Cape
survey'' in November 1997 (Ashoka et al. 2000, Martinez et al. 2001 and
references therein).  During the survey the CP stars HD\,13038, HD\,13079,
HD\,98851 and HD\,102480 have been discovered to be new $\delta$ Sct
pulsating variables, while HD\,12098 has been discovered to be a rapidly
oscillating Ap star. The pulsation properties of HD\,13038 (Martinez et al.
1999a), HD\,13079 (Martinez et al. 1999b) and HD\,12098 (Girish et al. 2001)
were reported earlier while those of HD\,98851 and HD\,102480 are presented
here.

Joshi et al. (2000) discovered two main periods of 75\,min and 150\,min for
HD\,98851. These pulsations were confirmed by Zhou (2001) with independent
observations taken from the Xinglong Station of the National Astronomical
Observatory, Chinese Academy of Sciences (NAOC).
The spectral
classifications of HD\,98851 and HD\,102480 are F1/F1IV/F3 and F2/F4/F4 on
the basis of K line, H line and metal lines, respectively (Abt 1984). These
classifications, as well as the presence of strong Sr\,II lines, weak Ca
K-line, and other indications of general abnormality confirm that HD\,98851
and HD\,102480 are marginal Am stars. Here we present the new high-speed
photometric and low-resolution spectroscopic observations of HD\,98851 and
HD\,102480. The basic physical parameters and colour indices of both stars are
given in Table 1. They are taken from Hauck \& Mermilliod (1998), Abt (1984)
and (ESA, 1997).

\section{Photometric Observations}

High-speed photometric observations of HD\,98851 and HD\,102480 were
obtained using a three-channel fast photometer (Ashoka et al. 2001, Gupta et
al. 2001) attached to the f/13 Cassegrain focus of the 104-cm Sampurnanand
telescope of the State Observatory, Naini Tal. All the observations from
Naini Tal were acquired as continuous 10-s integrations through
a Johnson $B$ filter. A photometric aperture of 30 arcsec was used to
minimize light losses due to seeing fluctuations and tracking errors. We used the
photometer in the single-channel mode as we were primarily searching for rapid
oscillations in CP stars in the period range of 4-16 min, which is typical
for rapidly oscillating Ap (roAp) stars. The observations were interrupted
for occasional sky background measurements to take account of changes in sky
brightness during the night. The observations reported here were carried out
during photometric sky conditions. While single-channel high-speed
photometry is not the ideal technique for studying the longer periods found
in $\delta$ Sct stars, under good photometric conditions it has been shown
to be adequate. Indeed, the light curves in Figs 1 and 2 show that.

HD\,98851 was also observed from NAOC, China, on two consecutive nights
using a three-channel fast photometer attached to the 85-cm Cassegrain
telescope. These observations comprised continuous 10-sec integrations
through a Johnson $V$ filter, also with an aperture of 30 arcsec. Unlike the
Naini Tal observations, these were simultaneous three-channel observations
of HD\,98851, a comparison star, SAO~62526, and the sky background.

Complete details of the photometric observations carried out
from Naini Tal, India, and NAOC, China, are listed in Table 2. The data are
available from the first author of this paper on request. The first
column of Table 2 lists the name of the star, the second column lists the UT
date on which the observations were obtained, the third column lists the
Heliocentric Julian Day (HJD) of the start of each run, the fourth column
gives the total duration of observations in hours and the fifth column lists
the total number of 10-s integrations obtained. The sixth column lists the
passband of the observations. The seventh and eighth columns list the frequency and
amplitude of the dominant oscillations detected in a given light curve, as
described in Section 4 below. The photometric observations of HD\,98851
taken from the NAOC are marked with asterisks; the rest are Naini Tal
observations. In total observations were obtained on 10 nights for
HD\,98851 and 6 nights for HD\,102480. Most of the observations spanned more
than one hour on a given night, thus covering at least one cycle of the variation.

\begin{table*}
\caption{Journal of observations of HD\,98851 and HD\,102480. On the two nights
marked with asterisks the data were taken from NAOC, China, while on other nights
 they were taken from Naini Tal, India.}
\begin{center}
\begin{tabular}{|c|c|c|c|r|c|r|r|}
\hline
Star & UT & HJD & Duration & Total~ & Filter & Freq.~~~ & Amp.~~~\\
     &Date & Start of the Run & (hr)&points & & (mHz)~~~ &(mmag)\\
\hline
 & & & & & & & \\
HD\,98851 & 24-01-2000 & 2451568.36543 & 0.81 & 294 & $B$ & 0.23 $\pm$ 0.15 & 23.1 \\
& & & & & & & \\
& 28-01-2000 & 2451572.38751 & 3.20 & 1155 & $B$ & 0.16 $\pm$ 0.07 & 17.7  \\
& & & & & & 0.06 $\pm$ 0.04 & 7.4 \\
& & & & & & & \\
& 17-02-2000 & 2451592.28338 & 3.88 & 1400 & $B$ & 0.20 $\pm$ 0.04 & 21.4 \\
& & & & & &  0.05 $\pm$ 0.03 & 11.5 \\
& & & & & & & \\
&19-02-2000& 2451594.19116 & 7.38 & 2660 & $B$ & 0.21 $\pm$ 0.02 & 21.9 \\
& & & & & & 0.10 $\pm$ 0.02 & 12.9 \\
& & & & & & 0.13 $\pm$ 0.02 & 4.1 \\
& & & & & & & \\
&21-02-2000 & 2451596.20508 & 6.90 & 2484 & $B$ & 0.21 $\pm$ 0.02 & 24.4 \\
& & & & & & 0.11 $\pm$ 0.02 & 12.5 \\
& & & & & & 0.03 $\pm$ 0.02 & 10.5 \\
& & & & & &  & \\
& 15-03-2000 & 2451619.24205 & 2.42 & 1169 & $B$ & 0.19 $\pm$ 0.07 & 11.9  \\
& & & & &&  0.07 $\pm$ 0.06 & 5.0 \\
& & & & &  & & \\
&18-03-2000 & 2451622.23641 & 3.33 & 1201 & $B$ & 0.06 $\pm$ 0.05 & 25.4 \\
& & & & &&  0.21 $\pm$ 0.06 & 21.1 \\
& & & & &&  0.11 $\pm$ 0.06 & 12.3 \\
& & & & & & & \\
&23-03-2000 & 2451627.16090 & 3.36 & 1212 & $B$ & 0.19 $\pm$ 0.05 & 27.9 \\
& & & & & & 0.12 $\pm$ 0.06 & 12.2 \\
& & & & & &  & \\
&13-04-2001$^*$ & 2452013.01264 & 6.68 & 2408 & $V$ & 0.11 $\pm$ 0.03 &16.8 \\
& & & & & & 0.06 $\pm$ 0.03 & 13.9 \\
& & & & & & 0.18 $\pm$ 0.02 &7.6 \\
& & & & & &  & \\
& 14-04-2001$^*$ & 2452014.00296 & 7.12 & 2564 & $V$ & 0.06 $\pm$ 0.05 & 12.4 \\
& & & & &&  0.20 $\pm$ 0.03 & 10.3 \\
& & & & &&  0.12 $\pm$ 0.02 & 7.6 \\
& & & & &  & & \\
\hline
& & & & & &  & \\
HD\,102480 & 19-01-2000 & 2451563.41189 & 0.95 & 343 & $B$ & 0.20 $\pm$ 0.04 & 30.5 \\
& & & & & & & \\
& 23-02-2000 & 2451598.33902 & 3.89 & 1400 & $B$ & 0.10 $\pm$ 0.04 & 23.0 \\
& & & & & & 0.19 $\pm$ 0.03 & 22.7 \\
& & & & & &  & \\
& 24-02-2000 & 2451599.19278 & 6.98 & 2684 & $B$ & 0.16 $\pm$ 0.02 & 15.9 \\
& & & & & & 0.10 $\pm$ 0.02 & 23.5  \\
& & & & & & 0.20 $\pm$ 0.03 & 10.0  \\
& & & & & & & \\
& 16-03-2000 & 2451620.22134 & 3.53 & 1269 & $B$ & 0.19 $\pm$ 0.04 & 19.3 \\
& & & & & & 0.07 $\pm$ 0.03 & 7.6 \\
& & & & & & 0.13 $\pm$ 0.03 & 5.0 \\
& & & & & &  & \\
& 27-02-2002 & 2452333.18516 & 6.25& 2249 & $B$ & 0.09 $\pm$ 0.04 & 12.1 \\
& & & & & &  0.19 $\pm$ 0.02 & 11.4 \\
& & & & & & & \\
& 03-03-2002 & 2452337.17280 & 7.30 & 2628 & $B$ & 0.10 $\pm$ 0.02 & 22.3 \\
& & & & & &  0.18 $\pm$ 0.02 & 16.0 \\

& & & & & & & \\
\hline
\end{tabular}
\end{center}
\end{table*}

\begin{figure}
\psfig{height=4.1in,width=3.64in,file=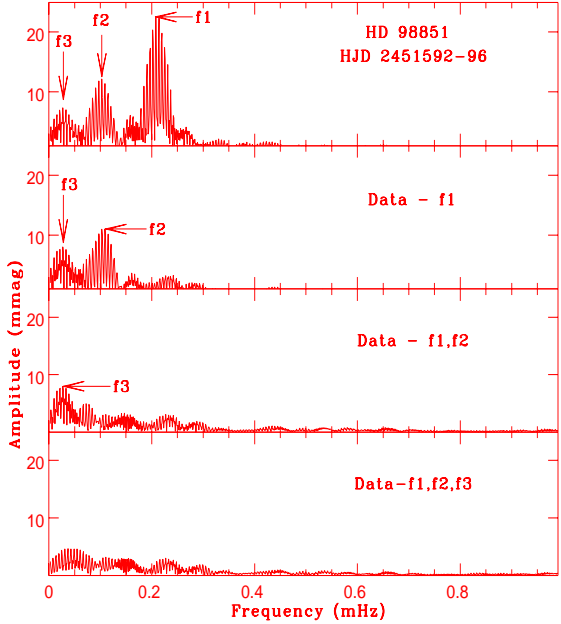}
\caption{Identification of the frequencies present in HD\,98851 for 
the combined data from nights JD 2451592--96, given in 
Table 2. The top panel shows the discrete Fourier transform (DFT) for the 
data set. The subsequent panels display the DFT after successive prewhitening 
of the frequencies $f_1$ to $f_3$ given in Table 3.}
\end{figure} 

\section {Photometric Data Reduction}

The photometric data reduction procedure was standard and consisted of the
following steps: dead-time correction to the count rates, subtraction of the
linearly interpolated sky background at the time of each observation and
extinction correction as a function of air mass. The HJD at the mid-point of
each integration was calculated to an accuracy of $10^{-5}$\,d
($\sim$1\,s). Some typical light curves of HD\,102480 are shown in Fig.
1 along with their amplitude spectra. The light curve obtained on HJD
2451598 is about 4 hr duration, while the light curve obtained on HJD
2452333 is around 6 hr duration. For both light curves it is clear 
that there is more than one frequency present in the data. In fact,
the alternating high and low amplitude cycles are indicative of the nearly-harmonic
relation of the two highest amplitude frequencies.

\begin{figure}
\psfig{height=4.1in,width=3.64in,file=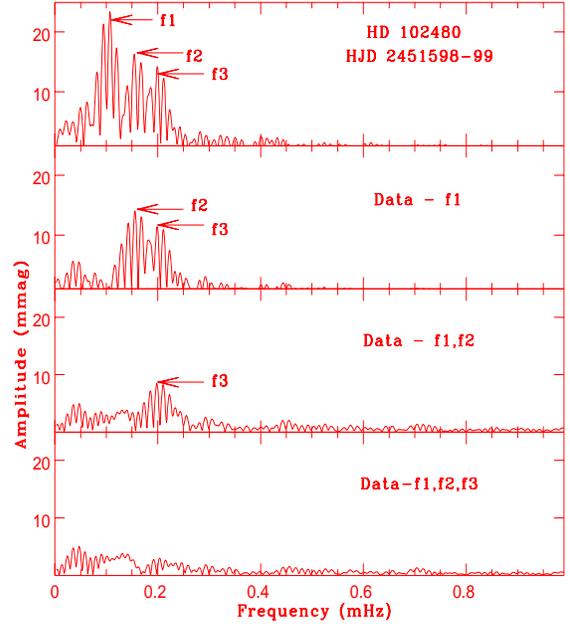}
\caption{Identification of the frequencies present in HD\,102480 for the combined data from nights JD 2451598--99, given in 
Table 2. The top panel shows the discrete Fourier transform (DFT) for the 
data set. The subsequent panels display the DFT after successive prewhitening 
of the frequencies $f_1$ to $f_3$ given in Table 3.}
\end{figure} 

Fig. 2 shows the light curves and the corresponding amplitude spectra of
HD\,98851 on three different nights. The light curves obtained on HJD
2451594 and HJD 2451596 are from Naini Tal, while that of HJD 2451014 is
from NAOC. The presence of amplitude modulation is clear in all the light
curves. The alternating high and low amplitude cycles are indicative of a
nearly-harmonic relation of the two highest amplitude frequencies. This is
very similar to the light curves seen for HD\,102480 in Fig. 1, but is not
at all typical of $\delta$ Sct stars in general.

\section {Frequency Analysis}

The time series data for both objects were analysed for periodic signals
using a fast algorithm (Kurtz 1985) based on Deeming's Discrete Fourier Transform (DFT)
for unequally spaced data (Deeming 1975). We adopted the criterion given by
Breger et al. (1993) that a peak in the amplitude spectrum is intrinsic when the signal
to noise (S/N) ratio is greater than 4; this can be lowered to 3.5 for
multi-periodic variables.

The first step of our analysis was to inspect the DFTs for each individual
light curve where the dominant frequency was identified. To identify other
frequencies present in the data, a sinusoid, $A_1 \cos(2\pi f_1t+\phi_1)$,
corresponding to the dominant frequency, amplitude and phase was subtracted
from the time series, a process we call `prewhitening.' The residuals to
this fit were then used to compute the DFT again, and the resulting dominant
frequency was identified as $f_2$. The prewhitening procedure was repeated
until the residuals were judged to be only noise, or the remaining
amplitude, even if real, was not identifiable with a particular frequency.
The frequencies and the corresponding amplitudes thus obtained are listed in
columns 7 and 8 of Table 2. We have listed here only the prominent
frequencies. All these Fourier transforms are ``raw'' transforms in the
sense that no low-frequency filtering to remove sky transparency
fluctuations was applied to the data, other than a correction for mean
extinction. Inspection of the nightly amplitude spectra for both stars shows
that all the signal power is contained in the range 0.0--0.5\,mHz.

We analysed the frequency content of data of both stars in a variety of ways,
but for the sake of clarity, we will discuss only the most instructive
analysis performed for each star. The data for each star were analysed in single nights,
in groups of closely spaced nights, and as a whole. Although the combined
full data sets had the highest frequency resolution, they also had the most
complex window patterns, because of the sampling gaps in the data. Figures 3
and 4 present the particular data subsets for HD\,98851 and HD\,102480 which
we found most instructive. 

\begin{figure}
\psfig{height=3.3in,width=3.5in,file=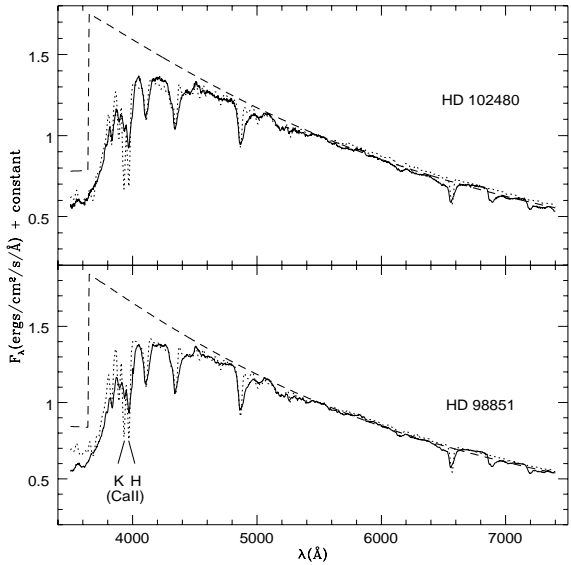}
\caption{The best matched synthetic spectra of Kurucz models, including continuum as well as lines, are overplotted with the observed spectra of the stars 
HD\,98851 and HD\,102480. The dotted and dashed lines represent line and continuum Kurucz models while the solid lines indicate our observed spectra.}
\end{figure}

Fig. 1 shows two typical light curves of HD\,102480, together with their
amplitude spectra. 
There is a clear
indication in these spectra of the presence of two main frequencies, $f_1=0.09$\,mHz and
$f_2=0.19$\,mHz, of varying amplitude. Fig. 2 shows three typical light curves for HD~98851 and their
amplitude spectra. The presence of at least two frequencies ($f_1 = 0.20$\,mHz
and $f_2 = 0.10$\,mHz) of variable amplitude is indicated. To define these
frequencies better, and to search for other frequencies present in the data,
we computed the DFTs of combined data sets for each star in groups of
consecutive, or closely spaced nights with long light curves. The iterative
prewhitening procedure described above was repeated until no more window
patterns could be identified above the level of the noise in the amplitude
spectrum.

Fig. 3 shows the analysis of 3 nights spanning 2451592--96 for HD~98851.
Three frequencies were identified in these data. The frequencies $f_1$ and
$f_2$ are real oscillations in the star, but our measurements of these frequencies suffer from a 1 d$^{-1}$ alias ambiguity. The frequency $f_3$ is {\em very} tentative because, at such low frequencies, we cannot discriminate between real oscillations in the star and sky transparency variations in these
single-channel data. We mention $f_3$ only because it appears in a number of
the nightly amplitude spectra listed in Table 2, and because we wish to draw
attention to the need for a differential photometric study of the variations
in HD~98851. Fig. 4 shows the analysis for the two nights 2451598--99 for
HD~102480. In this star, too, our frequency identifications suffer from 1 d$^{-1}$ alias ambiguities, particularly in the cases of $f_2$ and $f_3$.

The frequencies thus identified for both stars were fitted simultaneously to
the combined data by linear least-squares, which assumes that the frequencies
are perfectly known and adjusts the amplitudes and phases. The function
\begin{displaymath} B = A_{i} \cos ({2\pi f_{i}(t-t_0)+\phi_i}) \\
\end{displaymath}
was fitted by least squares, where $t_{0}$ was taken as the time of first
observation of the combined data set (Table 2). The fitted amplitudes and
phases, along with their errors, are listed in Table 3. We caution that the
frequencies listed in Table 3 are all subject to 1 d$^{-1}$ cycle count
ambiguities in the present study. A more extensive study would reduce these ambiguities, but to
eliminate them altogether, a multi-site campaign is necessary.

\begin{table*}
\caption{Simultaneous least-squares fits of the three most prominent frequencies fitted simultaneously to the
JD~2451592--96 data for HD~98851 (top) and the JD~2451598--99 data for
HD~102480 (bottom).}
\begin{tabular}{cccccc}
\hline
Star & Freq. & Amp. & $\phi$ & $t_0$\\
Data set & (mHz) & (mmag) & (rad) & (day)\\
&&&&& \\
\hline
&&&&& \\
HD~98851       &   0.208 $\pm$0.001  &   22.35 $\pm$ 0.12   &      -2.37 $\pm$ 0.01 & 2451592.28338\\
JD~2451592--96 &   0.103 $\pm$0.002  &   11.19 $\pm$ 0.12   &      -1.26 $\pm$ 0.01 & \\
               &   0.027 $\pm$0.002 &    8.03 $\pm$ 0.11   &      -1.79 $\pm$ 0.02 & \\
&&&&&\\
\hline
&&&&& \\ 
HD~102480      &   0.107 $\pm$0.002  &   20.77 $\pm$ 0.14   &       2.82 $\pm$ 0.01 & 2451598.33902\\
JD~2451598--99 &   0.156 $\pm$0.002 &   12.50 $\pm$ 0.14   &      -1.49 $\pm$ 0.01 & \\
               &   0.198 $\pm$0.003  &    9.28 $\pm$ 0.15   &      -3.08 $\pm$ 0.02 & \\
\hline
\end{tabular}
\end{table*} 

\begin{table*}
\caption{The physical parameters of HD\,98851 and HD\,102480 determined using
our spectroscopic data and Hipparcos data.}
\begin{center}
\begin{tabular}{ccccccccccccccc}
\hline
& && \\
Objects & $T_{eff}$ & log g & Spectral & & & & &Distance & $M_v$ & $M_{bol}$ & $
 log (L/L_{\odot})$  & $M/M_{\odot}$ & $R/R_{\odot}$ \\
Name & (K) & (cgs) & Type & & & & & (pc) & (mag) &(mag) & & & \\
& &&& & \\
\hline
& &&& & \\
HD\,98851 &7000$\pm$250 & 3.5$\pm$0.5 & F1IV & & & & & 171$\pm$25 &
1.2$\pm$0.3 & 1.1$\pm$0.3& 1.5$\pm$0.1 & 2.2$\pm$0.2 &3.6$\pm$0.6  \\
& &&& & \\
HD\,102480 &6750$\pm$250 & 3.0$\pm$0.5 & F3III/IV & & & & & 240$\pm$60 & 
1.5$\pm$0.5 & 1.4$\pm$0.5&1.4$\pm$0.2& 2.1$\pm$0.3 &3.5$\pm$0.9  \\
&&&& & \\
&&&& & \\
\hline
\end{tabular}
\end{center}
\end{table*}

\section{New Spectroscopic Observations and Data Reduction}

To measure the effective temperatures and surface gravities of HD\,98851 and
HD\,102480, we obtained low-resolution spectroscopic observations with the
104-cm Sampurnanand telescope (f/13) of the State Observatory, Naini Tal on
18 April 2002. We used a $1K \times 1K$ CCD detector and the HR-320
spectrograph, giving a linear dispersion of $\sim 2.4$\,\AA ~per pixel. The
spectra were taken using a 300-line\,mm$^{-1}$ grating and a 3-mm circular
aperture. We covered a spectral range of 3500--7400\,\AA. Apart from the
spectrophotometric standards we also observed two standard stars of similar
spectral type, HR\,5447 (F2V) and HD\,140283 (F5). 

The spectroscopic data
reductions were performed using the 
IRAF\footnote{IRAF is distributed by the National Optical Astronomy Observatories,
which are operated by the Association of Universities for Research
in Astronomy, Inc., under cooperative agreement with the National
Science Foundation.}
software package (Tody 1993). To estimate the accuracy of the 
spectral data we determined synthetic Johnson $V$ magnitudes for HD\,98851
and HD\,102480. Standard Kurucz (1993) models with solar metallicity and
micro-turbulent velocity of 2\,km\,s$^{-1}$ were used to match the
observed spectra by normalising the flux at 5500\,\AA. We considered our
spectra to have zero reddening. The best matched $T_{eff}$ and $\log g$
parameters for both stars are tabulated in Table 4. The best matched Kurucz
models in both lines and continuum are shown in Fig. 5 along with the
observed spectra. The $\log g$ estimates for both stars indicate that
they are giants or sub-giants of luminosity class III/IV. 

On comparing our spectra with
the observed stars of similar spectral type we estimate that HD\,98851 and
HD\,102480 are of spectral type F1 and F3, respectively. 
In Fig. 5 it is
notable that the Ca\,II H and K lines are weak compared to the Kurucz
models, supporting the Am classification of these stars. We also estimated
the equivalent spectral type by using the calibration of Schmidt-Kaler (1982) with the
absolute colour ($B-V$) and $ T_{eff}$. This comparison is consistent with our
spectral types of F1 and F3 for HD98851 and HD 102480, respectively, with
luminosity class III or IV for both stars.
Abt (1984) classified HD\,98851 and HD\,102480 with a Ca\,II K line type, Balmer line
type, metal line type as Am(F1/F1\,IV/F3) and Am(F2/F4/F4), respectively; our  spectral type estimates are in good agreement with his.

\section{Physical Parameters and Evolutionary Status of HD\,98851 and HD\,102480}

We derived several physical parameters for HD\,98851 and HD\,102480 using
Hipparcos data as well as from our observational data. A comparison of the
observed colours of HD\,99851 and HD\,102480 with those corresponding to the
spectral types determined in the last section indicates that both stars
suffer little interstellar reddening. The Hipparcos parallaxes of HD\,98851
and HD\,102480 are 5.84$\pm$0.87 mas and 4.17$\pm$1.04 mas, respectively.
The corresponding distances of these stars are $171\pm25$\,pc and
$240\pm60$\,pc. At these distances the absolute magnitudes of HD\,98851 and
HD\,102480 are $+1.23 \pm 0.32$ and $+1.50 \pm 0.54$, respectively, as given in
Table 4. Adopting a bolometric correction of -0.11 and -0.14 (Schmidt-Kaler
1982) for HD\,98851 and HD\,102480 respectively, we derived the
corresponding luminosities $\log (L/L_{\odot})$ as 1.12 and 1.36,
respectively. 

The masses for both the stars were calculated using the following 
empirical relation (Schmidt-Kaler 1982):
\begin{displaymath}
\log(M/M_{\odot}) = 0.46 - 0.1 M_{bol}.
\end{displaymath}
It is worth noting that this relation is determined from well-observed
eclipsing and visual binaries, and so does not depend on evolutionary
models.
The calculated masses of HD\,98851 and HD\,102480 are $2.2M_\odot$ and 
$2.1M_\odot$ respectively, which are well within the range of  
masses for $\delta$ Sct stars. The radii of the stars can be derived according to the radiation law (Schmidt-Kaler 1982):
\begin{displaymath}
\log({R/R_\odot}) = - 0.2 M_{bol} - 2 \log T_{eff} + 8.472,
\end{displaymath}
giving 3.6\,R$_\odot$ and 3.5\,R$_\odot$ for HD\,98851 and HD\,102480,
respectively. From the spectroscopic observations we determined the effective
temperatures of HD\,98851 and HD\,102480 to be 7000$\pm$250\,K and
6750$\pm$250\,K
respectively, while the corresponding $\log g$ values are 3.5$\pm$0.5 and
3.0$\pm$0.5, respectively. All the physical parameters derived are listed in Table 4.

\begin{figure}
\psfig{height=3.0in,width=3.5in,file=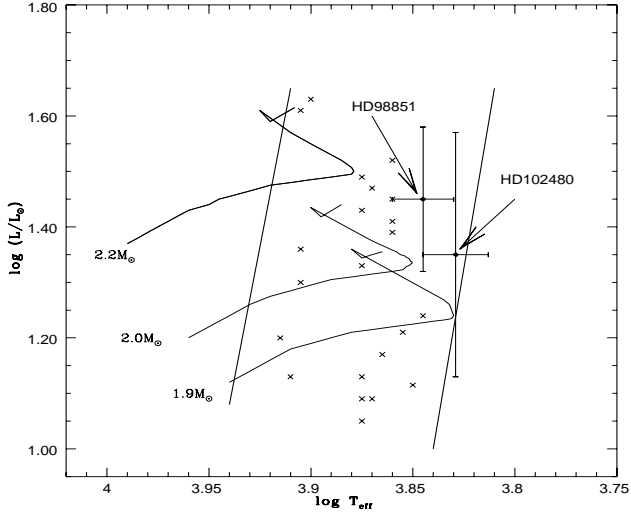}
\caption{An HR diagram showing the positions of HD\,98851 and HD\,102480 
in relation to the borders of the instability strip. The crosses indicate 
the positions of other known pulsating Am stars (Turcotte et al. 2000).
Three evolutionary model tracks for 1.9$M_\odot$, 2.0$M_\odot$, and
2.2$M_\odot$ stars are shown.}  
\end{figure} 

The positions of HD\,98851 and HD\,102480 in a theoretical HR
diagram are shown in Fig. 6, along with the positions of other known pulsating 
Am stars taken from Turcotte et al. (2000).
Stellar evolutionary model paths for $1.9M_\odot$,
$2.0M_\odot$ and $2.2M_\odot$ mass models are indicated in the 
figure where it 
is clearly seen that both HD\,98851 and HD\,102480
lie within the instability strip near the red edge.

Fig. 6 also indicates that both objects are probably near H core exhaustion,
but, within the errors they could be past it. The age of the model
corresponding the calculated luminosities and effective temperatures is
around 1.0 Gyr. Given their unusual pulsation characteristics these would be
interesting stars to model in more detail.

\section{Discussion} 
The unusual pulsations in the cool, marginal, evolved Am stars  
HD\,98851 and HD\,102480 
are very interesting. The alternating high and low amplitude variations with  
nearly harmonic (or sub-harmonic) frequency ratio close to 2:1 are 
conspicuous. In some $\delta$~Sct stars,  
a 2:1 frequency ratio is seen, and in others a pattern of alternating  
high and low amplitude peaks with  
longer periods is observed. The co-existence of both these phenomena 
has not previously been observed in chemically peculiar stars. 

Arentoft et al. (2001) found that XX Pyx shows a near 2:1 frequency ratio,
and suggested that 
XX Pyx 
is a $\delta$ Sct star in a binary system, in which the companion is  
low mass or compact object.  
A similar scenario cannot be excluded for HD\,98851 and HD\,102480, since
the spectroscopic and photometric evidence suggests that they are Am stars,
and it is known that Am stars are almost all in short-period known 
binary systems.
In the case of these two stars, radial
velocity observations are required to demonstrate their binarity. 
This, in turn, would constitute additional support for the Am
classification for these two stars.

The locations of HD\,98851 and HD\,102480 in the HR diagram 
are consistent with the theoretical work of Turcotte et al. (2000), which shows
that young Am stars are stable against $\kappa$-mechanism pulsation, and
that as they evolve towards the red edge of the instability strip they become 
unstable. However the ratio of two prominent periods for HD\,98851 and 
HD\,102480 has an intriguing value which is 
far from the ratio 0.75 to 0.79 associated with the radial fundamental 
and first overtone pulsation for a large sample of double-mode 
$\delta$ Sct stars.

Among the low-amplitude $\delta$ Sct stars which exhibit near 
2:1 frequency ratios are AN Lyn (Rodriguez et al. 
1997, Zhou 2002), V663 Cas (Mantegazza \& Poretti 1990) and 63 Her 
(Breger et al 1994). In all these cases the authors 
concluded that a mixture of radial and non-radial modes is needed to explain the 
non-standard frequency ratio. Therefore, in both of HD\,98851 and HD\,102480
there may be mixture  of radial and non-radial modes.

\section{Conclusions}

In this work we have presented the results of high-speed photometric and low
resolution spectroscopic observations of the Am stars HD\,98851 and HD\,102480.
Analyses of the available data show that HD\,102480 is pulsating
mainly in two frequencies, 0.107\,mHz and 0.198\,mHz corresponding to periods
2.6 hr and 1.4 hr. Similarly, HD\,98851 is pulsating mainly with two
frequencies 0.208\,mHz and 0.103\,mHz, corresponding to periods 1.34 hr and
2.70 hr. Beside the two main frequencies, we can see evidence of one other
frequency in the amplitude spectra of both stars.

The effective temperature and $\log g$ for the stars are determined to be
$7000 \pm 250$\,K, $3.5 \pm  0.5$ and $6750 \pm250$\,K, 
$3.0 \pm 0.5$ for HD\,98851
and HD\,102480, respectively.  The corresponding equivalent hydrogen line
spectral types are found to be F1IV and F3III/IV. 

We conclude that HD\,98851 and HD\,102480 are unusual variables lying
near the red edge of the instability strip. Their Am spectral types are
securely established; given their luminosities these stars belong to
the $\rho$ Pup group (previously known as $\delta$ Del) of evolved Am
stars. Their unusual nearly harmonic (or sub-harmonic) period ratios,
unusually high overtones and Am spectral types make these stars especially
interesting objects for further observational and theoretical studies.

\section*{Acknowledgments}
We are thankful to the referee for constructive suggestions
which improved the scientific content of the paper. 
This work was carried out under the Indo-South 
African Science and Technology Cooperation Programme joint project titled
``Naini Tal-Cape Survey for roAp stars,'' funded by the Indian and South 
African governments.

\end{document}